# The Inner Workings of Windows Security

Ashvini A. Kulshrestha

*Department of Computer Science*

*and Engineering*

*The Ohio State University*

*Columbus, OH, United States*

kulshrestha.18@buckeyemail.osu.edu

Guanqun Song

*Department of Computer Science*

*and Engineering*

*The Ohio State University*

*Columbus, OH, United States*

song.2107@osu.edu

Ting Zhu

Department of Computer Science

and Engineering

The Ohio State University

Columbus, OH, United States

zhu.3445@osu.edu

*Abstract*— The year 2022 saw a significant increase in Microsoft vulnerabilities, reaching an all-time high in the past decade. With new vulnerabilities constantly emerging, there is an urgent need for proactive approaches to harden systems and protect them from potential cyber threats. This project aims to investigate the vulnerabilities of the Windows Operating System and explore the effectiveness of key security features such as BitLocker, Microsoft Defender, and Windows Firewall in addressing these threats. To achieve this, various security threats are simulated in controlled environments using coded examples, allowing for a thorough evaluation of the security solutions' effectiveness. Based on the results, this study will provide recommendations for mitigation strategies to enhance system security and strengthen the protection provided by Windows' security features. By identifying potential weaknesses and areas of improvement in the Windows security infrastructure, this project will contribute to the development of more robust and resilient security solutions that can better safeguard systems against emerging cyber threats.

I. INTRODUCTION

As the predominant operating system for desktop and laptop PCs, Microsoft's Windows holds the biggest share of the market, at 69% [6]. This widespread usage makes it an attractive target for cybercriminals who aim to exploit vulnerabilities and deploy malware on these systems. Consequently, understanding and strengthening the security of Windows is crucial for ensuring the safety and privacy of users worldwide.

Windows security is not just confined to traditional desktop environments. With the advent of IoT [16-19] and smart applications [20-33], security considerations extend to a diverse ecosystem [34-39]. Windows systems must now consider the intricacies of low-power communication and leverage machine learning [40-48] to anticipate and counteract cyber threats. Furthermore, the convergence of Windows with other technologies like WiFi, BLE, and Zigbee [49-56], and innovative communication methods like backscatter, introduces new security challenges. However, this paper aims to explore how Windows security adapts to these evolving paradigms.

To address security concerns and protect users, Microsoft has implemented a range of security features in Windows, such as Windows Firewall, BitLocker, and Microsoft Defender. These features aim to provide comprehensive protection against various attack vectors and safeguard user data.

The objective of this project is to delve into the security features of Windows, explore common vulnerabilities, and validate their effectiveness using coding examples. Additionally, the project seeks to identify any vulnerabilities that may still exist despite the presence of these security features. By examining the security landscape of Windows, this research aims to provide insights and recommendations for enhancing the overall security of the platform, ultimately providing benefit to users and organizations that rely on Windows for their daily operations.

By focusing on both the strengths and weaknesses of Windows security, this project seeks to contribute to the ongoing conversation around securing this popular operating system. The findings and recommendations presented in this study should be valuable for IT professionals, security experts, and software developers seeking to improve the safety and resilience of Windows systems in the face of ever-evolving cyber threats.

II. BACKGROUND AND MOTIVATION

*A. Background*

According to the 2023 Microsoft Vulnerabilities Report, there were originally 333 vulnerabilities reported in 2023, and there has been an uphill trend in the years since. [7] While 2022 marks an all-time high, the trend started to plateau from 2020 onward. The most likely cause of this is the Microsoft customers helping improve Windows, which helps Microsoft patch vulnerabilities out of the system. A robust, yet scalable systematic approach to counter such vulnerabilities is thus a necessity. For instance, while BitLocker has a mild impact on performance, by encrypting the entire hard disk the data becomes much harder for various threats to read, thus protecting the data from attack both in real-time and when at rest. Tan et al [1] analyzed the security of BitLocker from a variety of angles, recommended countermeasures, and suggested security enhancements for BitLocker encryption. Furthermore, various authors [2] analyzed vulnerabilities in the latest Windows operating systems' default settings, which allow attacks on the computers from malware on USB drives.

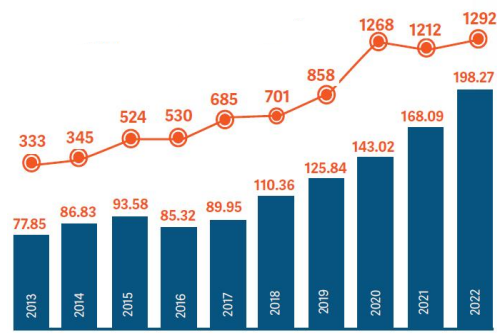

Figure 1 Microsoft's Annual Revenue Worldwide vs. Total Vulnerablities (2013-2022) [7]

*B. Motivation*

Many high-profile attacks have been launched against Windows systems. One such attack, NotPetya used multiple vectors and used an encryption process to destroy data at blinding speeds, causing over $10 billion in total damage. [4] One company that was hit by the attack, Maersk, lost over $300 million despite only having one device that was vulnerable to NotPetya. Conficker is another dangerous threat that was first known in 2008, but still caused comparable damage at roughly $9.1 billion. [8,9] It infects and then remotely controls systems to perform a variety of actions, such as causing denial of service, system crashes, spreading other forms of malware, and outright shutting down essential services like banks, hospitals, and government offices. [9]

As there are other notorious attacks such as WannaCry, Stuxnet, and the attack against Solarwinds in 2020, the main takeaway from their impact is that our systems need to be properly hardened and defended against threats like them. While Windows has taken action by implementing defensive systems against them, there may still be holes in their security.

### III. PROPOSED APPROACH

The proposed approach consists of several stages aimed at evaluating the security features of Windows, identifying common threats, and providing recommendations to enhance system security. These steps are as follows:

1. Research: Investigate the key security features of Windows, including BitLocker, Microsoft Defender, and Windows Firewall, to gain a comprehensive understanding of their capabilities and limitations.
2. Threat Identification: Identify common security threats targeting Windows systems, and evaluate the potential risks they pose to the security of these systems based on existing knowledge and documented cases.
3. Effectiveness Evaluation: Assess the effectiveness of Windows security features against the identified threats, focusing on their ability to prevent, detect, and mitigate these threats.
4. Improvement Recommendations: In cases where the security features fail to completely address the identified threats, research and propose solutions to enhance their effectiveness. Additionally, document any potential risks or drawbacks associated with the proposed solutions, ensuring that the recommendations are balanced and well-informed.

By following this approach, the project aims to provide valuable insights into the current state of Windows security, identify areas for improvement, and contribute to the ongoing efforts to safeguard Windows systems against ever-evolving cyber threats.

### IV. REVIEW OF WINDOWS KEY SECURITY FEATURES

As a part of this project, I was focused on Windows Defender, Firewall and BitLocker. A big benefit of these features is that they are included with every copy of Windows, so no further monetary cost is necessary. Proper configuration of these features would improve security posture against various threats. While the basic features are already able to protect against a wide variety of threats, using different settings may be necessary to protect against the threats that pose the biggest danger to the Windows systems, especially because certain threats may target certain vulnerabilities precisely because the default settings have said vulnerabilities since that would be a common denominator against most systems.

*A. Microsoft Defender*

Microsoft Defender is a built-in feature of Windows systems that comes with many security features, such as cloud backup storage and a strong firewall. It supports threat and vulnerability management, attach-surface reduction, endpoint detection and response, and automatic investigation and remediation. However, there are also other important features that Microsoft Defender lacks, so novel threats can be designed to avoid it. Some of these missing important features include VPN support and a password manager, and Defender struggles to detect zero-day threats. Zero-day threats specifically are a huge threat regardless of whether a system is properly built to intercept them because once they enter that system, they act faster than the target system could respond, giving the end user "zero days" to deal with them. Thus, it would be too late to do anything about a newly started zero-day attack. Microsoft Defender is a built-in feature of Windows systems and can operate on the cloud.

*B. BitLocker*

BitLocker is a full-volume encryption feature included with Microsoft Windows versions starting with Windows Vista [5], with the purpose of data protection. While activating BitLocker results in slightly decreased read/write performance, there are many desirable features that have utility far outweighing this drawback, so it is a very secure solution. Since BitLocker is a part of Windows, it does not come with any additional licensing costs. It also has other benefits in the use of Trusted Platform Module (TPM), which carries out cryptographic operations, and end users being able to automatically save keys to an Active Directory. One of Windows 11's system requirements is a TPM 2.0, so that need for BitLocker may already be satisfied.

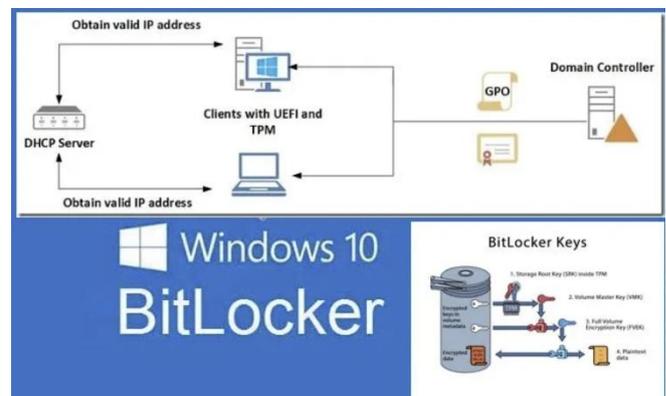

Figure 2: BitLocker Pin bypass unlock configuration.(Image credit: https://techdirectarchive.com/2021/01/31/bitlocker-pin-bypass-how-to-configure-network-unlock/)

BitLocker comes with BitLocker Drive Encryption, a data protection feature that integrates with the operating system and addresses the threats of data theft or exposure from computers left inaccessible by bad actors.

*C. Windows Firewall*

Windows Firewall has been a feature of Windows systems since 2004 and is very easy to set up. [13] However, Windows Firewall's basic protection does not provide the same protection that more expensive firewalls do, and experienced hackers can either find routes around it or disable it via malware.

## V. EXPERIMENTS & STEPS COMPLETED

A comprehensive approach was taken to plan various experiments.

*A. Environment Setup*

I started with one dedicated Windows laptop where I installed Hyper-V software so that I can create a VM for targeting attacks. While I wanted to have two VMs installed so that I can simulate more tests, it became quickly evident that with my current laptop configuration of 16GB RAM and 1TB SSD, that performance of laptop was terrible with 2 VMs. Hence, I setup just one VM and setup another test machine in case I need two target machines instead of 2 VMs. After setting up Jupyter Notebook and other softwares, I was ready for simulating cyberattacks on target VM. The main machine has various modules installed that can launch cyberattacks against the virtual machine. The command *ipconfig* was run on the virtual machine to obtain its IP address, after which Nmap was used from the main machine to check whether the virtual machine had any accessible ports on the 1-32767 range. After that, the main machine launched attacks via Metasploit on the virtual machine, and any response from the virtual machine's Microsoft Defender was logged.

The virtual machine had Windows Defender turned on by default. Code was written for checking its status and turning it on if it was disabled but turning it off via "MpCmdRun.exe -wddisable" to see if the code would turn it back on has yet to result in a successful attempt. Looking at the logs for attempting this command, the most likely explanation for this command's failure is that the –wddisable argument for the MpCmdRun.exe command is forbidden. In the PowerShell commands, running MpCmdRun.exe with an incoherent argument (for instance, "MpCmdRun.exe -jfhbjlfdkbhlpqrieuo") returned a different error code than –wddisable, which supports this theory.

*B. Initial Experimentation*

Using Python, I wrote a script to check the status of Windows Defender. If it was off, the script goes ahead and turns it own. Although simple in nature, the approach here is that one could programmatically check key features automatically and alert users to leverage inbuilt security features within Windows Security to provide a robust configuration for patching up more vulnerabilities.

*C. Metasploit*

When using Metasploit, I first installed Metasploit framework. An exclusion needed to be provided to Defender as specified by the metasploit website and the installer needed to be run by running command prompt as administrator and executing it through that, but the process was otherwise somewhat straightforward when those are accounted for. I used Metasploit to simulate security threats and monitored the vulnerabilities of the security solutions.

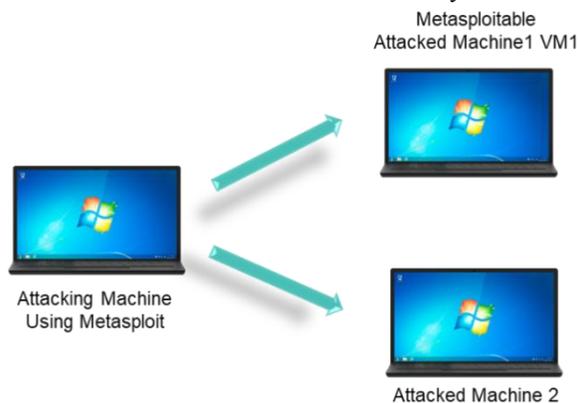

Figure 3: Setting up Testing Environment

I have written sample code in Python and in a .rc file to simulate an attack targeted at the VM and monitor whether Windows Security catches it. My code implements booting up and shutting down a VM, as well as obtaining its IP address. It can also perform port scans, which can help find open ports on the VM.

I found that Microsoft Defender protected it against all of those attacks that I created against the target machine. I have also made sure it was active on the VM when I was testing, but it could not be turned off.

The first step in this test was to run a search to identify which kind of threat would be applicable to our test scenarios. This can be done writing the command "Search Exploit" We can search one for Windows in particular. We can eliminate ones that do not apply as our target system may not have those software components.

I used windows/smb/ms17_010eternalblue. Even after finding an open port on the virtual machine, it was found that the target was not vulnerable. I made .rc files to have a set of commands as above so that appropriate sets of threat simulations can be invoked. During this testing, I also checked many other exploits such as kill-bill for a low-ranked case, various Oracle-related attacks, and some that relate to Microsoft Office, but it was found that none of them were compatible either.

While there are tools available (for cost) to monitor and analyze such simulated attacks, during my experiment I mostly reviewed the messages displayed at the console to see if an exploit was able to reach the target. From my experience, it seems the firewall intercepted most of the attacks. I monitored various system resources to ensure if any exploits were successful in damaging the system, but it turns out that none were.

During the course of this project, further experiments were conducted to determine whether Nmap scans could reach other ports based on certain security features of the virtual machine. Additionally, various tools, including Scapy,

OWASP ZAP, and OpenVAS, were employed to simulate network-based attacks against the virtual machine, providing a comprehensive assessment of its security posture.

### D. Nmap

First, I installed the python-nmap library using pip. I used Nmap to simulate security threats and took note of whatever vulnerabilities of the security solutions were found. I have written sample code in Python to simulate an attack targeted at the VM and monitor whether Windows Security catches it. My code implements booting up and shutting down a VM, and in addition to that can extract its IP and perform port scans. If the IP address of the virtual machine is made known, a thorough port scan can be run to potentially expose one of its ports for targeting.

I further looked into examples for Firewall/IDS Evasion and Spoofing using Nmap framework. Given the nature of such operations, I have to ensure that I am running all the commands as an administrator.

I watched process maps and error messages from Windows Defender to make sure simulated attacks were being trapped by Windows security and appropriate settings were enabled.

### E. Scapy

Scapy is a Python-based packet manipulation tool, library, and interactive shell that allows for craft and analysis of packets. It can be used to create custom network protocols and security tests. Plenty of examples for them can be found at https://scapy.readthedocs.io/en/latest/, which I have carried out to get a feel for the program.

After installing Scapy on my main machine, I simulated various key scenarios to sniff network traffic.
Its purpose is to layer various network protocols within packets and send them to whatever destinations the end user has in mind. When these packets find their targets, they either return to the machine that sent them or come with a premade "answer" from the destination. Scapy supports the ability to track the geographic path a packet takes en route to its destination through the various stops it makes along the way. If I develop a means of reaching my virtual machine with Scapy, it should hopefully expose a vulnerability for me to exploit on that virtual machine. After that, I shall run Metasploit and target the virtual machine's vulnerability to conduct testing on the limits of its Microsoft Defender system. Scapy also supports the development of new types of packets, such as layering existing types of packets on old ones. It can send packets to, or "blanket ping" a certain set of hosts to see which ones respond; Sending a packet to "192.168.1.5\30" for instance sends packets to IP addresses 192.168.1.4 through 192.168.1.7 and takes note of which ones reached their targets. I got even more experience with scapy through following the "Scapy in 0x30 minutes" tutorial. During the traceback section, I tried to replace the destination IP address with my target VM's IP address, but no packet I sent returned.

### F. Exploit Database (Exploit-DB):

Exploit Database is a comprehensive database of exploits, vulnerabilities, and shellcodes that can be used as a resource for creating custom security tests. One could search for exploits and vulnerabilities related to specific systems or applications, and then develop custom tests based on the resulting findings. It has a searchsploit app that can be downloaded on Kali Linux systems and run to research much like Metasploit can be downloaded on Kali Linux systems, though there does not seem to be support for it on Windows-based systems right now. Clicking each of the entries in ExploitDB will provide relevant information on the exploit itself, with entries typically having not only the code needed to execute them, but also documentation and relevant links on the exploits themselves. If the operating system family that the target uses is known, then the database can be browsed for entries that target it.

### G. OWASP Zed Attack Proxy (ZAP):

OWASP ZAP is an open-source web application security scanner that not only focuses on web applications but also offers features for testing the security of web servers and underlying systems. ZAP provides an API that enables users to interact with the tool programmatically, including using Python scripts, making it a versatile solution for security testing.

During the course of this project, I encountered some challenges in setting up a compatible Java Runtime Environment (JRE) to run ZAP. After overcoming these initial obstacles, I decided to run ZAP against the URL "https://localhost:8080", representing the host system itself. Although many of the issues ZAP identifies are primarily related to web-based systems and servers, the tool also provides remediation solutions for the detected vulnerabilities, outlining the respective deployment phases for applying these fixes when appropriate.

Subsequently, I replaced the URL with the virtual machine's port to target it using ZAP. However, the "handshake" required for ZAP to scan the target was rejected. This observation led to the conclusion that rejected handshakes are likely the reason why Metasploit and Scapy were also unable to reach the virtual machine, as handshakes are essential for two devices to exchange information.

This valuable insight into the role of handshakes in security testing has highlighted the importance of ensuring that security tools can establish proper communication with the target system. To enhance the effectiveness of security testing, it is essential to address handshake-related issues and ensure that the chosen tools can successfully communicate with and analyze the targeted systems. By doing so, organizations can gain a more comprehensive understanding of their security posture and better protect their systems from potential cyber threats.

### H. Automated Tools for Security Testing-OpenVAS

OpenVAS is an open-source vulnerability scanner that allows an end user to automate the process of discovering and managing vulnerabilities in their network and systems. It includes a large database of vulnerability tests and supports scripting languages like Python for automation. However, in researching means of trying to run OpenVAS on a Windows system it was discovered that opening it on a

Linux virtual machine on a Windows system was the only way to run it. [14] This unfortunately meant that I was unable to explore OpenVAS to the best of my ability as I am focused on Windows security and did not have a Linux machine. Below is my research on this tool set and I have added another tool set Nessus as the replacement tool under section I.

OpenVAS enables a more detailed analysis of the virtual machine's vulnerabilities was possible, allowing for the identification of potential weaknesses and the development of appropriate mitigation strategies. The integration of OpenVAS with other security testing tools, such as Scapy and OWASP ZAP, enabled a multi-layered approach to security testing that increased the likelihood of discovering potential vulnerabilities within the system.

*I. Nessus*

Nessus is a vulnerability scanner designed and owned by Tenable. While a seven-day free trial is available, it costs thousands of dollars to use yearly and initialization is a lengthy process for it. Even after that, compiling plugins will be a lengthy process too, but only one in three million scans is a defect, resulting in staggering accuracy. [15] Still, from the results, Nessus was able to resolve the names of both machines. It was able to uncover a vulnerability of 5.3 on the main machine but had only managed to uncover information about the virtual machine itself and not any vulnerabilities thereof, showing that the virtual machine is protected well from outside attacks.

## VI. ISSUES ENCOUNTERED

Over the course of this project, several challenges were faced while investigating the security features of Windows, testing their effectiveness against common threats, and proposing recommendations for improvement. Some of the key issues encountered are as follows:

1. Limited Access to Vulnerability Information: Due to the sensitive nature of security vulnerabilities, detailed information about certain vulnerabilities and exploits might be restricted or unavailable to the public. This limitation posed challenges in understanding the complete threat landscape and evaluating the effectiveness of Windows security features. Additionally, Microsoft being a proprietary platform offered limited insights to their internal architecture.
2. Complex Security Features: Windows security features, such as BitLocker, Microsoft Defender, and Windows Firewall, have evolved over time and become increasingly sophisticated. Understanding their intricacies and the various configuration options available required a significant amount of research and technical knowledge.
3. Dynamic Threat Landscape: The constantly evolving nature of cybersecurity threats makes it challenging to stay updated on the latest attack vectors, exploits, and malware. The project required continuous monitoring of new developments in the field to ensure that the analysis and recommendations remained relevant.
4. Testing Limitations: Simulating real-world attack scenarios in a controlled environment proved to be challenging, as it required access to specialized tools and resources. Additionally, ethical considerations had to be taken into account while conducting security tests to avoid causing harm to systems or violating legal restrictions.
5. False Positives and Negatives: While testing the effectiveness of Windows security features, the project encountered instances of false positives (i.e., benign activities being flagged as malicious) and false negatives (i.e., actual threats going undetected). Distinguishing between these occurrences and actual security incidents required careful analysis and verification.
6. VM Setup: Creating the VMs had several issues, though these were resolved by research into the limitations of RAM requirements. Additionally, setting them up in the first place was very time-consuming.
7. Software compatibility: One of the proposed tools, OpenVAS, was found to be solely available on Linux after some research, but I was able to find another tool that was equally applicable.
8. Local network responses: One of the programs I ran inadvertently shut off the printer at my home, but simply restarting the printer solved that issue.

Despite these challenges, the project managed to provide valuable insights into the current state of Windows security features, identify potential vulnerabilities, and propose recommendations for enhancing system security. However, it is essential to acknowledge the limitations and uncertainties that arose during the project due to the issues encountered.

## VII. OBSERVATIONS AND CONCLUSIONS

The experiments conducted in this study provided valuable insights into the effectiveness of Windows Defender and other integrated security features in defending against various simulated cyberattacks. The main observations and conclusions drawn from these experiments are as follows:

- Robustness of Windows security features: The inability to successfully compromise the target virtual machine demonstrates that Windows Defender and other built-in security solutions are capable of providing robust protection against a range of potential threats.
- Importance of proper configuration: The experiments highlighted the critical role that proper security configuration plays in mitigating potential attacks. Ensuring that security features are enabled and correctly configured can significantly improve the overall defense against cyber threats.
- Handshake rejection as a defense mechanism: The difficulty in targeting the virtual machine, such as the rejection of handshakes, underscores the effectiveness of certain security mechanisms in preventing unauthorized access to the system.

These mechanisms can be crucial in safeguarding a system from potential attacks.
- Limitations of existing security testing tools: The experiments revealed that some security testing tools and methods might not be fully effective in targeting more recent or up-to-date operating systems. This highlights the need for continued development and improvement of security testing tools to keep pace with the evolving threat landscape.
- Significance of a multi-layered security approach: The experiments emphasized the importance of employing a combination of security solutions, such as firewalls, intrusion detection systems, and antivirus software, to provide comprehensive protection against cyber threats.

Overall, the experiments conducted in this study demonstrated that the Windows security features, when properly configured and used in conjunction with other security solutions, can provide a strong defense against a variety of cyberattacks. However, it is essential to continue refining and updating these security features to stay ahead of emerging threats and maintain a high level of protection.

## VIII. Future work

There are several directions for future work that can be explored, with a focus on expanding and building upon the experiments already conducted:
- Expanding the range of simulated attacks: Conduct experiments with a wider variety of exploits, including more recent vulnerabilities, to further test the capabilities of Windows security features and identify potential areas for improvement.
- Investigating the impact of different security configurations: Examine how varying security settings and configurations within Windows can impact the effectiveness of the security solutions in defending against cyber threats.
- Exploring advanced evasion techniques: Investigate how more sophisticated evasion techniques, such as traffic obfuscation and packet fragmentation, can bypass or challenge the defenses provided by Windows security features.
- Developing custom security tests: Create custom security tests based on the findings from the conducted experiments to further explore potential vulnerabilities and weaknesses in the Windows security features.
- Analyzing the impact of machine learning and artificial intelligence on security defenses: Investigate how incorporating machine learning and AI into security solutions can enhance their ability to detect, respond to, and predict potential threats.
- Investigating integration with other security solutions: Examine how Windows security features can be integrated with third-party security solutions to provide a more comprehensive defense against cyber threats.
- Assessing the effectiveness of Windows security features in cloud environments: Explore how Windows security solutions perform in cloud-based environments and identify potential areas for improvement in securing virtual machines and cloud infrastructure.

By pursuing these directions for future work, a more comprehensive understanding of the strengths and weaknesses of Windows security features can be developed, informing the design of more effective security solutions and strategies to protect against evolving cyber threats.

In addition, Future research should also focus on the implications of evolving wireless communication technologies [49-56] and their impact on Windows security frameworks. This includes the challenges of securing backscatter communication and ensuring interoperability with various wireless standards. Additionally, exploring the application of Windows security in smart health and sustainable energy systems [56-64] can reveal new insights into how Windows can be fortified in diverse application domains. Moreover, the integration of Windows with cloud computing and microservices architectures [65-70] poses unique challenges for maintaining robust security in distributed systems, which we can further explore in this direction in the future